\PassOptionsToPackage{table}{xcolor}
\PassOptionsToPackage{bookmarks=false}{hyperref}
\documentclass[sigconf]{acmart}
\pdfoutput=1
\usepackage{etex}
\usepackage[utf8]{inputenc}

\copyrightyear{2018}
\acmYear{2018}
\setcopyright{licensedothergov}
\acmConference[ICSE-SEIP '18]{40th International Conference on Software Engineering: Software Engineering in Practice Track}{May 27-June 3 2018}{Gothenburg, Sweden}
\acmBooktitle{ICSE-SEIP '18: 40th International Conference on Software Engineering: Software Engineering in Practice Track, May 27-June 3 2018, Gothenburg, Sweden}
\acmPrice{15.00}
\acmDOI{10.1145/3183519.3183540}
\acmISBN{978-1-4503-5659-6/18/05}

\usepackage[table]{xcolor}
\usepackage{booktabs} 
\usepackage{numprint}
\usepackage{xspace}
\usepackage{cellspace}
\newcommand{\TODO}[1]{\textcolor{red}{todo: #1}}\newcommand\todo\TODO
\newcommand{\TODOSU}[1]{\textcolor{green}{todo: SU: #1}}\newcommand\todosu\TODOSU
\usepackage{listings}

\newcommand\nbfailures{\numprint{11523}\xspace}
\newcommand\nbprojects{\numprint{1609}\xspace}
\newcommand\nbrecommandations{\numprint{7}\xspace}
\newcommand\nbpatches{\numprint{15}\xspace}
\newcommand{\ra}[1]{\renewcommand{\arraystretch}{#1}}
\lstdefinelanguage{diff}{
  language=java,
  basicstyle=\ttfamily\scriptsize,
  sensitive=true,
  numbers=none,
  morecomment=[f][\color{gray}][0]{diff},
  morecomment=[f][\color{gray}][0]{index},
  morecomment=[f][\color{blue}][0]{@@},
  morecomment=[f][\color{magenta}][0]{***},
  morecomment=[f][\color{violet}][0]{!},
  morecomment=[f][\color{red!60!black}][0]{-},
  morecomment=[f][\color{green!60!black}][0]{+},
  morecomment=[f][\color{magenta}][0]{---},
  morecomment=[f][\color{magenta}][0]{+++},
  morecomment=[f][\color{gray}][0]{Binary},
  morecomment=[f][\color{gray}][0]{Only},
  morecomment=[f][\color{gray}][0]{old},
  morecomment=[f][\color{gray}][0]{new},
  morecomment=[f][\color{gray}][0]{rename},
  morecomment=[f][\color{gray}][0]{similarity},
  morecomment=[f][\color{gray}][0]{deleted},
  morecomment=[f][\color{magenta}][0]{***************},
  morecomment=[f][\color{red!60!black}][0]<,
  morecomment=[f][\color{green!60!black}][0]>,
  morecomment=[f][\color{blue}][0]{0},
  morecomment=[f][\color{blue}][0]{1},
  morecomment=[f][\color{blue}][0]{2},
  morecomment=[f][\color{blue}][0]{3},
  morecomment=[f][\color{blue}][0]{4},
  morecomment=[f][\color{blue}][0]{5},
  morecomment=[f][\color{blue}][0]{6},
  morecomment=[f][\color{blue}][0]{7},
  morecomment=[f][\color{blue}][0]{8},
  morecomment=[f][\color{blue}][0]{9},
}[comments]

\title{How to Design a Program Repair Bot?\\Insights from the Repairnator Project}
\author{Simon Urli}
\affiliation{
  \institution{University of Lille \& Inria Lille, France}
}
\email{simon.urli@inria.fr}

\author{Zhongxing Yu}
\affiliation{
  \institution{University of Lille \& Inria Lille, France}
}
\email{zhongxing.yu@inria.fr}

\author{Lionel Seinturier}
\affiliation{
  \institution{University of Lille \& Inria Lille, France}
}
\email{lionel.seinturier@inria.fr}

\author{Martin Monperrus}
\affiliation{
  \institution{KTH Royal Institute of Technology, Sweden}
}
\email{martin.monperrus@csc.kth.se}

\begin{abstract}
Program repair research has made tremendous progress over the last few years, and software development bots are now being invented to help developers gain productivity.
In this paper, we investigate the concept of a ``program repair bot'' and present Repairnator. The Repairnator bot is an autonomous agent that constantly monitors test failures, reproduces bugs, and runs program repair tools against each reproduced bug. 
If a patch is found, Repairnator bot reports it to the developers. 
At the time of writing, Repairnator uses three different program repair systems and has been operating since February 2017. 
In total, it has studied \nbfailures test failures over \nbprojects open-source software projects hosted on GitHub, and has generated patches for \nbpatches different bugs.
Over months, we hit a number of hard technical challenges and had to make various design and engineering decisions. 
This gives us a unique experience in this area. 
In this paper, we reflect upon Repairnator in order to share this knowledge with the automatic program repair community.
\end{abstract}

\begin{document}

\maketitle
\section{Introduction}
Program repair research has made tremendous progress over the last few years \cite{genprog,semfix,Xuan2017,prophet}. 
In a variety of empirical evaluations \cite{gensis,defects4j-repair,Angelix}, it has been shown that current program repair systems are able to synthesize patches for real bugs in real large programs. 
However, previous evaluations of program repair techniques generally only evaluate the capability of the repair algorithms themselves. 
For the use of program repair techniques in practice, several other phases such as failure detection, bug reproduction, and patch reporting are also needed before or after the run of the core repair algorithm itself.
To demonstrate the real potential of program repair in industry, it is desirable to study the design and implementation of an end-to-end repair toolchain that is amenable to the mainstream development practices.

For bridging this gap between research and industrial use, we investigate the concept of a ``program repair bot'' in this paper. 
To us, a program repair bot is an autonomous agent that constantly monitors test failures, reproduces bugs, and runs program repair tools against each reproduced bug. 
If a patch is found, the program repair bot reports it to the developers. 
We envision that in ten years from now there will be hundreds of program repair bots that will work in concert with developers to maintain large code bases. 
But today, to the best of our knowledge, nobody has ever reported on the design and operation of such a repair bot.

The Repairnator project is a project to design, implement and operate a repair bot for Java programs. 
This repair bot itself is called after the project: ``Repairnator'', and the name will only refer to the bot in the remaining of the paper.
Repairnator constantly monitors test failures happening in continuous integration, also called CI:
CI is a popular development practice  \cite{CIpopular1,CIpopular2} that involves frequently integrating and testing code changes.
When a test failure happens on CI, Repairnator first tries to reproduce the CI failure, then runs publicly available program repair tools to make a ``repair attempt'', and finally collects and reports information about the failure reproduction and repair attempt.

At the time of writing, Repairnator uses three different program repair systems and has been operating since February 2017. 
In total, it has studied \nbfailures test failures over \nbprojects open-source software projects hosted on GitHub, and has generated patches for \nbpatches different bugs. None of those patches were proposed to the developers because they all suffer from the overfitting problem \cite{qi2015efficient,smith2015cure,defects4j-repair}: they indeed fix the failing build but cannot be considered as a general, appropriate solution to the bug.

The design and operation of Repairnator has been challenging. 
Over months, we hit a number of hard technical challenges and had to make various design and engineering decisions. 
This gives us a unique experience in this area. 
In this paper, we reflect upon Repairnator in order to share this knowledge with the automatic program repair community.
For sake of open-science, all the data discussed in this paper is available at the following URL: \url{https://github.com/Spirals-Team/icse-seip-2018-repairnator}. 

The pipeline of Repairnator is constituted by three stages: CI Build Analysis, Bug Reproduction, and Patch Synthesis.
For each of the three stages: 
\emph{(1)} we present how it has been designed, aiming at inspiring the authors of upcoming repair bots; 
\emph{(2)} we report on results about the operation of Repairnator itself over 11 months of experiment; and
\emph{(3)} we present and discuss actionable recommendations on how to design a program repair bot based on our experience in architecting and operating Repairnator. 

To sum up, our contributions are:
\begin{itemize}
\item a blueprint design of a program repair bot for continuous integration (CI) build failures;
\item a set of unique empirical facts about program repair and bug reproduction collected over \nbfailures CI build failures across \nbprojects software projects;
\item \nbrecommandations actionable recommendations to help authors of future program repair bots.

\end{itemize}

The rest of this paper is organized as follows.
In Section~\ref{sec:overview}, we present an overview of Repairnator and the terminology used in this paper.
The three next sections are dedicated to present and discuss the different stages of the Repairnator approach. 
In Section~\ref{sec:stage_scanning}, we present the process of selecting projects and analyzing CI builds to repair.
We then discuss in Section~\ref{sec:stage_reproduction} the Repairnator's approach to reproduce a failing build. 
Section~\ref{sec:stage_repairing} presents the approach used to synthesize patches and the obtained results.
Related work about software development bots and program repair are given in Section~\ref{sec:related}, which is followed by conclusion remarks in Section~\ref{sec:conclusion}.

\section{Overview of Repairnator}
\label{sec:overview}

The Repairnator project is a project to make key scientific progress in the area of program repair. In particular, the Repairnator project consists of designing, implementing and running Repairnator.
Repairnator aims to propose patches for bugs to open-source developers before they have themselves fixed those bugs.
In other terms, \emph{Repairnator aims at being faster than humans to fix bugs}.

\subsection{Terminology}
Repairnator is designed for modern development toolchains based on continuous integration (CI), versioning with Git, and collaboration within development platforms such as GitHub.
We first define the key terminology related to this toolchain.

In a typical Git-like versioning process, every change is a commit. Optionally, branches are used to separate work made on new features, bug fixes, etc.
A continuous integration service (\emph{CI}) typically compiles the code and runs the tests for each commit made on any branch of the source code.
CI produces a \emph{build} for each change.
A build contains the information about the source code snapshot used (e.g. a reference to a Git commit), the result of CI execution (e.g. fail or success), and an execution trace or \emph{log}. Additional information such as code coverage may also be provided.

The execution of a build is triggered on different events: for example when a commit is pushed to a Git repository, or when a \emph{pull request} is proposed by a developer.
``Pull Request'' (PR) is a concept popularized by GitHub which consists in a change in the code proposed by an external developer of the project.
Pull requests are the main mechanism to encourage external contributors on open-source projects.
Pull requests can be examined and discussed by the maintainers of a project, who can ask for changes. Upon acceptance, the PR code is integrated to the main code base. A PR is said closed when it has been accepted or definitely refuse.

We also discuss in this paper the concept of ``merge commit'': this is a commit created by the merge of two different branches.
Merge commits are created when integrating in the main branch development with new changes.  
In the context of pull requests merge commits are very important because they are automatically created when a PR is accepted.

We have designed Repairnator to specifically work on top of a CI service:
Repairnator proposes patches in response to faulty CI builds (builds with status ``fail''). 
Those patches ``repair'' the commit that triggered the faulty build.

\subsection{Repairnator Workflow}

\begin{figure}
\centering
\includegraphics[width=1.0\columnwidth]{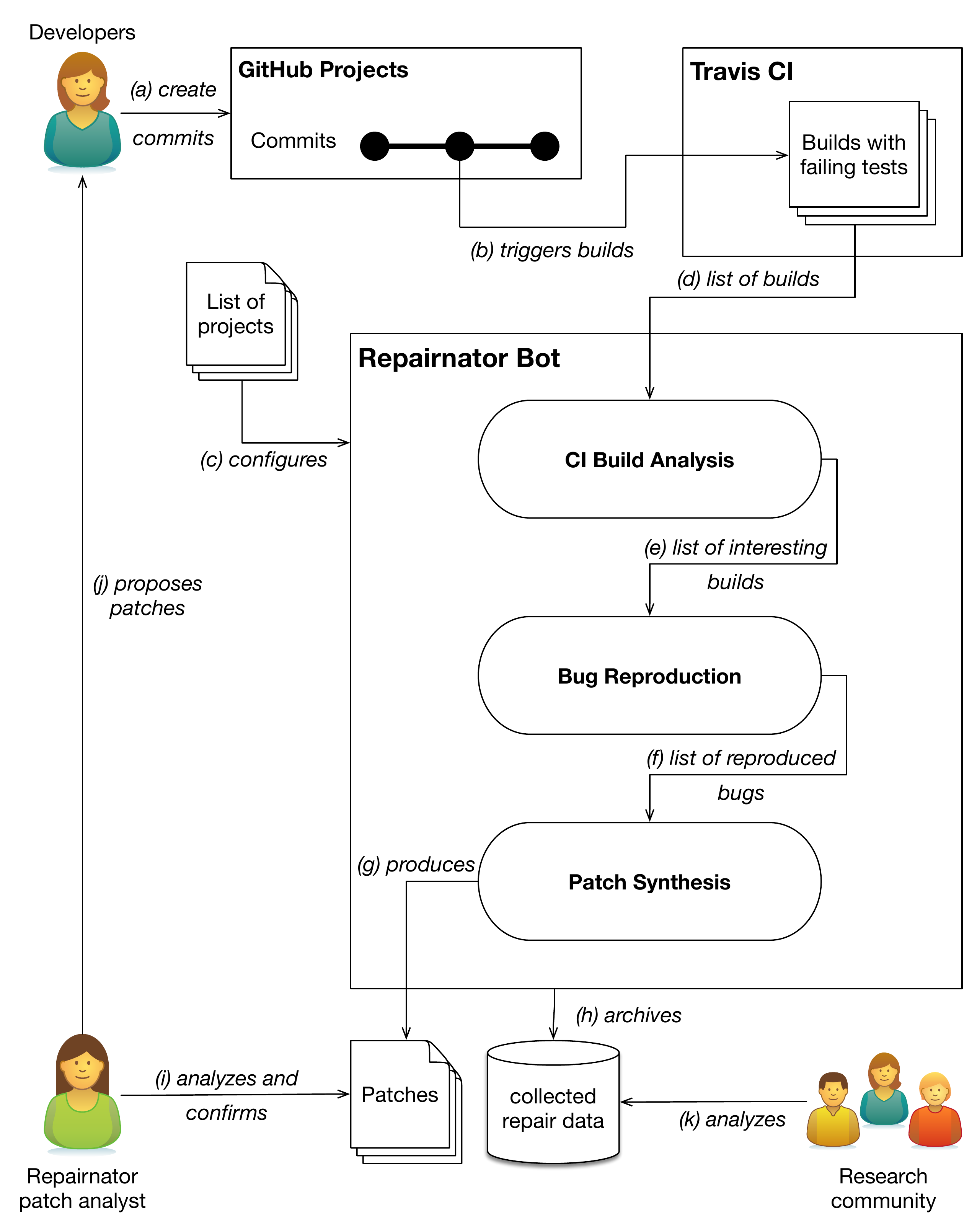}
\caption{The overview workflow of the Repairnator program repair bot.}
\label{fig:repairnator-overview}
\end{figure}

Figure~\ref{fig:repairnator-overview} gives an overview of how the Repairnator bot works.
The primary input of Repairnator are continuous-integration builds, triggered by commits made by developers (top part of the figure, arrows \emph{(a)} and \emph{(b)}).
The outputs of Repairnator are two-fold:
\emph{(1)} it automatically produces patches for repairing failing builds \emph{(g)}, if any;
\emph{(2)} it collects valuable data on program repair in the field \emph{(h)}, for future research in this area \emph{(k)}.

The Repairnator bot itself works as follows.
Continuously, it monitors all CI activity of projects coming from a specific configuration list \emph{(c)}.
The CI builds are given as input to a pipeline that contains three stages: 
\emph{(1)} a first stage, called CI Build Analysis, that collects and analyzes CI builds \emph{(d)} from GitHub projects \emph{(a and b)} (see Section~\ref{sec:stage_scanning});
\emph{(2)} a second stage, called Bug Reproduction, that aims at reproducing the build failures that have happened on CI (see Section~\ref{sec:stage_reproduction}); 
\emph{(3)} a third stage, called Patch Synthesis, that uses the failure reproduction information to search for patches (see Section~\ref{sec:stage_repairing}). 

As shown in Figure~\ref{fig:repairnator-overview}, the produced patches are first analyzed by a Repairnator patch analyst \emph{(i)}: if she finds a patch correct, she will then proposes it for the original developers of the project to fix the bug \emph{(j)}. As already mentioned, Repairnator also produces research data \emph{(k)} which can later be used by the research community. 

\subsection{Main Design Choices}
We present in this section the main design choices we made in the design of our repair bot.

\paragraph{GitHub and TravisCI} 
We want to plug Repairnator into serious large scale software projects.
There are many such projects in open-source.
GitHub is currently the largest open-source code hosting service in the world with 25.3 million  repositories active between September 2016 and September 2017~\cite{GitHub2017}.
Its ecosystem notably includes TravisCI, a state-of-the-art CI service, free for usage by open-source projects, providing an API to get access to build information.
We design the Repairnator bot to be integrated into the GitHub ecosystem to easily access a lot of open-source projects and all the related development information (commits, CI builds, activity, etc.).

\paragraph{Java projects} 
We design Repairnator to repair Java programs.
The main reason is that we have extensive expertise with automatic repair of Java software, and there are publicly-available state-of-the-art repair tools  that can be integrated in Repairnator including NPEFix~\cite{Cornu2015}, Nopol~\cite{Xuan2017} and Astor~\cite{Martinez2007}.
Since Java is the third most popular language on GitHub~\cite{GitHub2017}, it is easy to get a lot of active Java projects on GitHub.

\paragraph{Build-based repairing bot} Repairnator is meant to be used as part of the CI, which leads us to design it as a build-based tool.
Repairnator requires a build on a CI to be triggered.
There are different reasons for a CI build to fail, such as build script error, compilation error or test failure. Repairnator only focuses on the latter, i.e. it only performs test-suite based repair. All repair tools mentioned above  are indeed test-suite based repair tools.

\section{CI Build Analysis}
\label{sec:stage_scanning}

The first stage of Repairnator is to determine which software projects and CI builds are candidates to program repair. 
We present first the approach, then our key results obtained, and finally we discuss them. 

\subsection{Approach}
The goal of this stage is to detect and analyze failing builds in order to  prepare the repair attempts coming afterwards.
To achieve this, we devise an approach that involves two steps: the first step consists of choosing  interesting GitHub projects to be considered in the context of automated program repair.
This step is executed only once and the obtained list of projects is used for each execution of Repairnator (see Figure~\ref{fig:repairnator-overview} \emph{(c)}).
The second step aims at analyzing builds from the chosen projects to produce a list of interesting builds to repair (see Figure~\ref{fig:repairnator-overview} \emph{(e)}). 
This step is executed at the beginning of each Repairnator execution (also called Repairnator run).

\subsubsection{List of Considered GitHub Projects}
\label{sec:selecting-project}
We first define the following criteria for selecting projects:
(1) the project is open-source and available on GitHub;
(2) the project has a test suite;
(3) the project uses the Java language and the Maven building tool: our toolchain is dedicated to Java and Maven offers a build process centered on tests;
(4) the project is popular and active: we define this by having ``stars''\footnote{https://help.github.com/articles/about-stars/} on GitHub (a popularity note), and having commits or pull requests in recent history (activity in late 2016 in our case);
(5) the project uses the TravisCI continuous integration service, which is well integrated into the GitHub infrastructure and provides an API to get build information and results. 

At least two existing databases contain relevant information about GitHub projects: GHTorrent~\cite{Gousios2013} and TravisTorrent~\cite{Beller}.
We used them in order to produce a list of projects according to our criteria, without having to crawl GitHub ourselves.
TravisTorrent was a first choice to be able to use the third criteria concerning test suite information, but it contains less repository entries than GHTorrent.

Consequently, we used a first request on TravisTorrent and
we then used a more generic request on GHTorrent database using criteria 1, 2 and 4 in order to gather a large list of GitHub projects.
We finally filtered this list to only keep projects complying with criteria 3 and 5 by querying TravisCI and analyzing build logs.

\subsubsection{Analyzing Build Information}
\label{sec:create-build-list}

For each Repairnator execution, Repairnator collects all recent CI builds from the projects under consideration. It then outputs a list of failing builds subject to repair attempts.

In Repairnator, a build is considered as interesting to repair if it fulfills the following criteria: 
(1) it must be a failing build according to continuous integration (failing can mean several things: not compilable or tests not passing); 
(2) it must have at least one failing test;
(3) it must have finished after a specific date because we are only  interested in repairing fresh build failures.

The first criteria is checked with TravisCI API: it gives detailed information and metadata on a build. 
However, there is no direct information given by TravisCI API about test failures, 
consequently Repairnator parses build logs in order to assess the second  criteria.

In order to propose patches on the fly for the most recent failing builds, Repairnator is designed to be executed every few hours.
During our experiments, we mostly launch its execution every 4 hours. 
Hence, for each execution, Repairnator only considers builds that were completed within the last 4 hours.
This mechanism guarantees us that we respect the third criteria.

\subsection{Results}
We have operated Repairnator since January 2017 but the first month was a pilot experiment.
In this paper, all presented data span the period between February 1, 2017 and January 1, 2018.

\begin{figure}
\centering
\includegraphics[width=1.0\columnwidth]{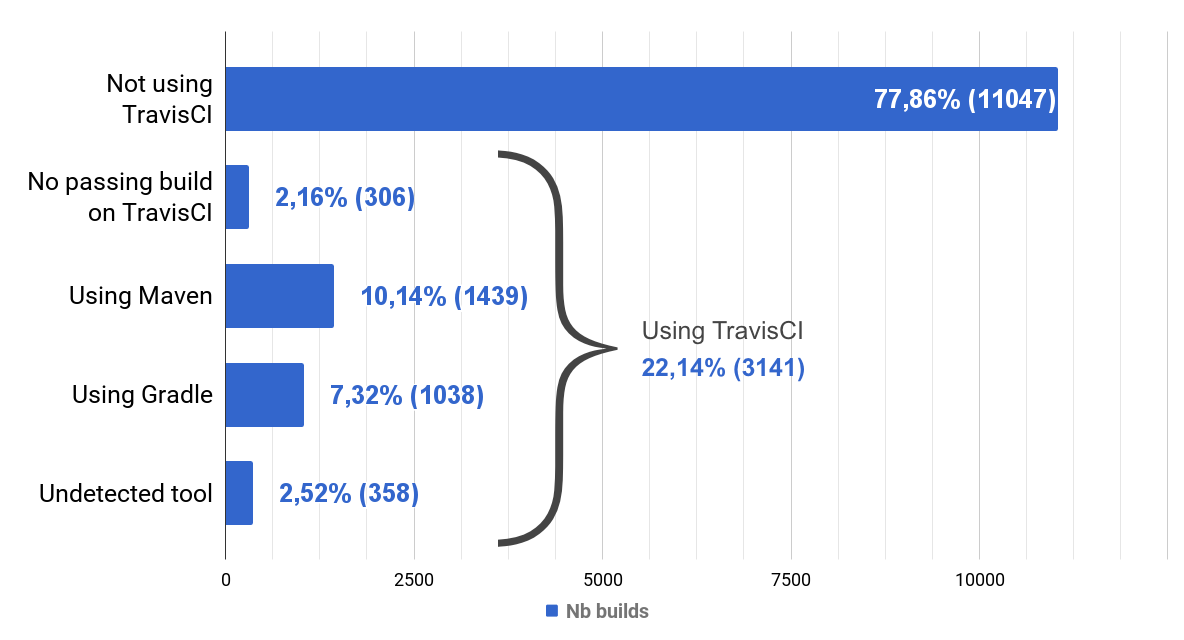}
\caption{The usage of tools over 14,188 Java projects hosted on GitHub.}
\label{fig:tools-repartition}
\end{figure}
We show in Figure~\ref{fig:tools-repartition} the distribution of build tools for \numprint{14188} popular Java projects we obtained on GHTorrent.
For this set of projects, we can see that 22.14\% of them (3141 projects) are using TravisCI.
However, 306 projects (2.16\%) do not have a single passing build on TravisCI: it usually means that the developers have not correctly set up the TravisCI configuration.

On the remaining projects using actively TravisCI, \numprint{1439} (10.14\%) use the Maven build tool and 1038 (7.32\%) use Gradle.
For 358 projects (2.52\%) we did not manage to identify the build tool automatically: they may use a build tool like Ant or Ivy.
Those numbers show that the focus on Maven is justified because this is where there are the most interesting builds to consider for repair.

\begin{figure}[h]
\centering
\includegraphics[width=1.0\columnwidth]{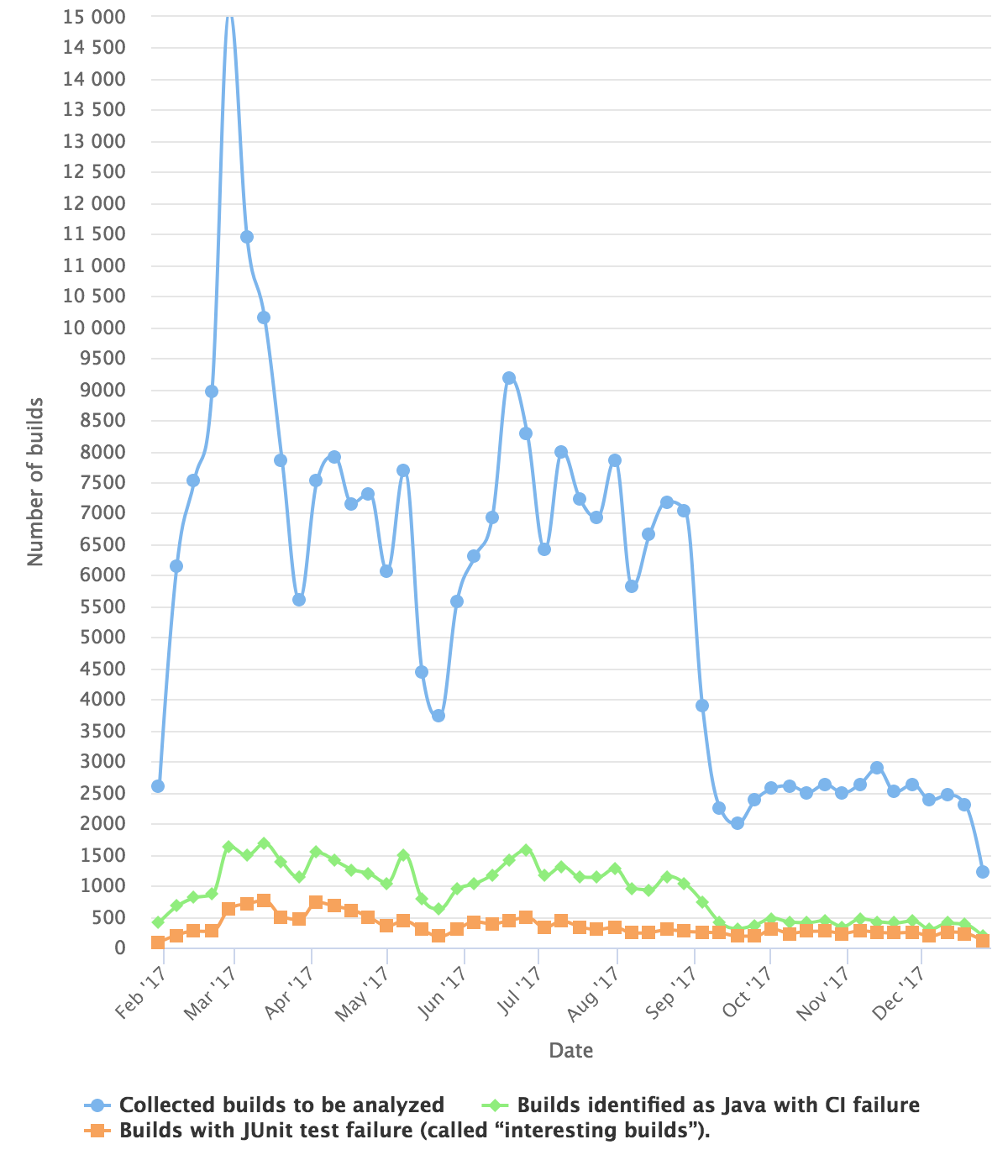}
\caption{Collected and analyzed builds number by operating Repairnator from February to January 2018.}
\label{fig:scanned-builds-timechart}
\end{figure}

We show in Figure~\ref{fig:scanned-builds-timechart} the number of collected builds per week from February 2017 to January 2018. 
This figure gives 3 different chart series: the blue one with circle marks represents the total number of collected builds to be analyzed; the green one with diamond marks represents the total number of Java builds with CI failure; and the orange one with square marks represents the total number of interesting builds, meaning the CI failing builds with identified JUnit test failures.

We can see that the overall number of interesting builds is considerably low compared to the overall number of collected builds.
The main reason is that most CI builds are successful.
We also see that on average half the number of CI failing builds are identified as containing test failures.
The other CI failing builds can be related to compilation error, checkstyle errors or code coverage errors, etc.
This shows the need to filter CI failing builds before trying to reproduce them.

We present in Table~\ref{tab:most-reproduced-projects} the 10 most active projects by the number of builds, with the total number of failing builds with failing tests as detected by Repairnator, and among those builds the ones coming from a GitHub pull request. 
For instance, there were 1000 builds with test failures on \emph{prestodb/presto}, this means that Repairnator has done 1000 repair attempts for it.
The number of build failures can be related to external activity:
as of January 2018, the \emph{prestodb/presto} Java project reports \numprint{2360} forks on its GitHub account, against 528 forks for the \emph{jabref/jabref} Java Project: this can explain the difference of activity (builds and pull requests) between the two projects.

We can also see that around 50\% of builds detected by Repairnator come from pull requests: this shows the importance of considering pull-request based development for program repair bots.
However, we notice the exception of \emph{evolveum/midpoint} project which has really few pull requests, but a high number of detected builds: this can be explained by a CI configured but not used by the developers. 
For this project, we observe  on the GitHub page of the project that almost all commits are linked to a failing build on TravisCI: the developers don't take into account the CI for fixing their project.

\begin{table}
\centering
\ra{1.3}
\rowcolors{2}{gray!20}{white}
\begin{tabular}{p{25mm}rrr}\toprule
Project & Builds with & PR builds with & \% of PR\\
& tests failures & tests failures & builds\\ \hline
prestodb/presto & 1000 & 889 &  88.90\% \\
druid-io/druid & 579 & 464 &  80.14\% \\
apache/flink & 477 & 349 &  73.17\% \\
apache/nifi & 472 & 327 &  69.28\% \\
hubspot/singularity & 437 & 114 &  26.09\% \\
apache/storm & 349 & 255 &  73.07\% \\
corfudb/corfudb & 313 & 151 & 48.24\% \\
spring-projects/spring-boot & 277 & 92 &  33.21\% \\
apache/zeppelin & 210 & 47 &  22.38\% \\
jabref/jabref & 201 & 82 & 40.80\% \\  \hline
Total on 468 projects & \numprint{11523} & \numprint{5874} & 50.98\% \\ 
Average on 468 projects & 25 & 13 & 50.98\% \\
\bottomrule
\end{tabular}
\caption{Projects with the most failing builds (between February and January 2018). Failing builds in the context of pull -request is notable.}
\label{tab:most-reproduced-projects}
\end{table}

\subsection{Recommendations}

We discuss here the recommendations for creating a list of projects and analyzing builds in a repair bot.
All recommendations presented here will be applied in the future version of Repairnator.

\subsubsection{List of Considered GitHub Projects}

Public databases of projects (e.g. TravisTorrent or GHTorrent) can be used to quickly gather set of projects filtered by some criteria (e.g. number of stars, number of pull requests, languages, etc).
However, what matters most for a build-based repair bot, is the actual project activity, which may not be reflected in the database metadata.
Requesting development platforms API (TravisCI API and GitHub API in our case) allows to get more data than contained in the project database, and also provides more up-to-date data.

\textbf{Recommendation \#1:} Check directly against the API from CI and code hosting services for building an appropriate up-to-date list of projects.

\subsubsection{Analyzing Build Information}

Analyzing build information becomes necessary with a large list of projects, when one wants quick results without replicating every single build because a replication attempt is resource consuming.
Filtering builds by looking on CI information like the status, build date, or duration of the build is easy and fast thanks to the CI API, and is enough for most usages.
When one wants a sharper filtering, parsing the logs of a build remains a possibility.
However, we strongly discourage this practice as it is too error prone.

\textbf{Recommendation \#2:} To the maximum extent, stick to the metadata provided by the considered CI service, and think twice before parsing logs, which is very tedious and error-prone.

\section{Local bug reproduction}
\label{sec:stage_reproduction}

This stage aims at locally reproducing a build failure observed in the continuous integration service.
We present the Repairnator approach, the important results we obtained, and then we discuss the approach. 

\subsection{Approach}
This stage takes as input the list of interesting failing builds previously computed (as presented in~Section \ref{sec:create-build-list}) and produces some test information like the number of running and failing test cases, their names, the elapsing time and the exception thrown by the failing test cases (see Figure~\ref{fig:repairnator-overview} \emph{(e)} and \emph{(f)}). 

It consists in the following steps:
\begin{enumerate}
\item checking out the code from GitHub;
\item compiling the code;
\item executing the tests;
\item observing the test outcomes.
\end{enumerate}

\subsubsection{Checking out the code}
The purpose of the first step is to get exactly the same source code as the failing build.
In the easy case, this only consists in cloning the repository from GitHub, getting a commit identifier of the failing build, and checking it out from the git repository.
However in the case of a pull request on GitHub, an additional step is required.
TravisCI performs a merge commit between the master branch and the PR branch before the start of a new build.
In order to get the same source code as TravisCI, Repairnator also performs this additional merge commit.
Finally, in some cases, commits might have been deleted: this happens for instance with amended commits (or branched updates with push-force option).
In such cases, the bug reproduction phase is stopped because the checking out of the code fails.

\begin{figure}
\centering
\includegraphics[width=1.0\columnwidth]{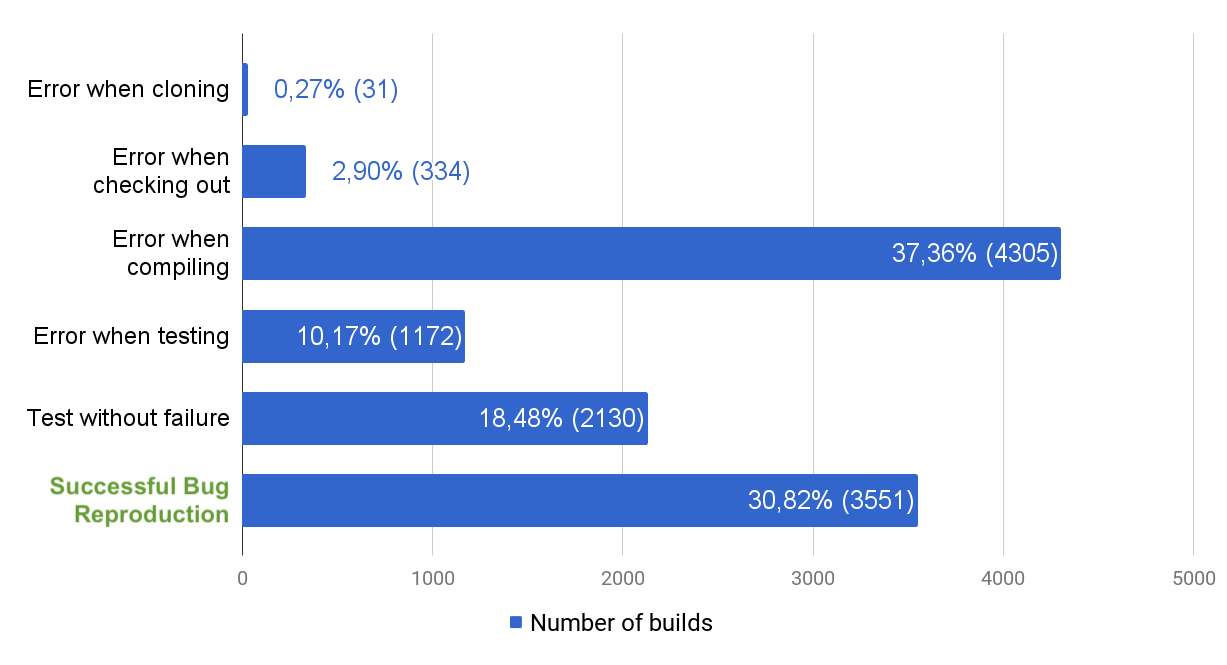}
\caption{The errors happening during local bug reproduction over \nbfailures reproduction attempts between February and January 2018.}
\label{fig:bug-reproduction-status}
\end{figure}

\subsubsection{Compiling the code}
This step consists in checking that the source code and the test code can be compiled. 
As Repairnator is designed to use Maven, this stage is performed by a call to the command \texttt{mvn install}.
This command first resolves the dependencies, compiles the source code of the application, compiles and runs the tests, packages it, and copies it in a local cache.
In order to avoid conflicting versions of the same dependencies, Repairnator creates a new Maven cache directory for each bug reproduction attempt.
Many projects use additional build checks (e.g. checkstyle, code coverage, etc.). Repairnator explicitly skips them because they are not relevant for the test-suite based repair tools we use.

\subsubsection{Executing the tests}
This step consists in executing all test cases of the checked-out version of the project.
The execution of the step consists in calling the command \texttt{mvn test}.
Note that certain Maven testing plugins have to be skipped for sake of consistency, like Maven Failsafe that aims at running integration tests: those tests usually need some specific resources to be run (e.g. an access to a database), which can not be reproduced in the Repairnator environment.

\subsubsection{Observing test outcomes}

This step consists in getting the test outcomes and observing if there are failing tests.
In practice, tests are run using standard test drivers such as JUnit or TestNG. The build tool itself (such as Maven Surefire) provides some abstraction over the test outcomes.
In Repairnator, we observe the test outcomes based on the API of Maven and JUnit.

\subsection{Results}

Figure~\ref{fig:bug-reproduction-status} shows the breakdown of bug reproduction errors.
Between February and January 2018, we have tried to reproduce \nbfailures failing builds. 
Repairnator has been successful at reproducing \numprint{3551} of them (30.82\%).
This shows the great difficulty of reproducing a bug locally: most software projects require a complex test environment that is really hard to locally mimic. This difficulty comes from the huge number of interacting tools involved in a typical build (much beyond the presence of Maven and the JVM in the context of Java projects), but also due to the presence of a myriad of versions. 

We identify 5 main causes for not reproducing a bug, corresponding to the different steps of the bug reproduction process.

The first cause is an error occurring during project compilation: \numprint{4305} bugs reproduction stopped at this point (37.36\%). 
This happens for different reasons such as (1) the usage of a dependency only available by manual installation or (2) the requirement of a specific version of Java.

The second cause is that we did not manage to reproduce a test failure: it happens for \numprint{2130} builds (18.48\%).
The reason can be the usage of a different environment compared to the CI, like a different version of the JDK.
We also notice that ``flaky'' tests are another issue: a ``flaky'' test is a test that passes or fails under the same circumstances leading to non-deterministic results.
A test might fail on CI due to flakiness, and passes locally by coincidence.

The third cause is about errors during the testing step (errors of the test harness itself, not of the application tests): it concerns \numprint{1172} builds (10.17\%).
There could be different reasons for those errors: the usage of another unit test framework than JUnit, or the execution of a specific Maven plugin.
For instance, some plugins are indeed configured to fail under some circumstances (e.g. when a threshold is not reached in code coverage).

The fourth cause concerns errors that occurs during the checking out step (getting the right commit): this happens for 334 builds (2.90\%), often because of a deleted commit.

Finally, for 31 builds, we did not manage to clone the repository (0.27\%): we had an issue managing our disk space when executing many bug reproductions at the same time, also, from time-to-time GitHub may have been unreachable from our network.

We show in Table~\ref{tab:most-reproduced-bugs} the 10 projects with the most reproduced bugs.
For instance, Repairnator was able to locally reproduce 359 build failures out of 579 CI failures (62\%) for project druid-io/druid.
We note that if \emph{prestodb/presto} project is the project with the most detected failing builds, we can see that we manage to locally reproduce bugs for only 19.40\% of its builds.
On the contrary, \emph{xethortio/jedis} is a project with a low number of detected failing builds, but we manage to reproduce 94\% of its builds.
These two tables show that the reproducibility of builds is really dependent of the projects under study.

\begin{table}[h]
\centering
\ra{1.3}
\rowcolors{2}{gray!20}{white}
\begin{tabular}{p{25mm}rrr} \toprule
Project & Builds with & Reproduced & \% of\\
& test failure & bugs & Reproduced\\ \hline
druid-io/druid & 579 & 359 & 62.00\% \\
apache/flink & 477 & 326 & 68.34\% \\
prestodb/presto & 1000 & 194 & 19.40\% \\
hubspot/singularity & 437 & 182 & 41.65\% \\
corfudb/corfudb & 313 & 126 & 40.26\% \\
apache/storm & 349 & 111 & 31.81\% \\
geoserver/geoserver & 118 & 109 & 92.37\% \\
spotify/docker-client & 111 & 99 & 89.19\% \\
xetorthio/jedis & 100 & 94 & 94.00\% \\
4pr0n/ripme & 94 & 87 & 92.55\% \\ \hline
Total on 468 projects & \numprint{11523} & \numprint{3551} & 30.82\% \\
Average on 468 projects & 25 & 8 & 30.75\% \\ \bottomrule
\end{tabular}
\caption{The ten projects with the most failing builds locally reproduced by Repairnator.}
\label{tab:most-reproduced-bugs}
\vspace{-17px}
\end{table}

Finally, Table~\ref{tab:most-common-failures} shows the 10 most commons test failure types we gathered.
\texttt{AssertionError} is the most common failure type, with more than \numprint{2100} occurrences observed.
In JUnit, \texttt{AssertionError} means that an assertion has failed, which is the essence of test-suite based repair: this validates the appropriateness of a CI repair bot such as Repairnator.
Immediately after, we can see that \texttt{NullPointerException} is an extremely frequent failure type, which confirms the relevance of program repair specific to memory errors.

We can see that these 10 failure types constitutes almost 70\% of the total number of failure types encountered (\numprint{4452} failure types against a total of \numprint{6380}).
This low diversity of failure types shows that targeting specific kinds of failure types can be an interesting strategy for designing future and effective repair tools.

\begin{table}
\centering
\ra{1.3}
\rowcolors{2}{gray!20}{white}
\begin{tabular}{p{50mm}r} \toprule
Exception & Occurrences \\ \hline 
java.lang.AssertionError & \numprint{2162} \\
java.lang.NullPointerException & 641 \\
org.junit.ComparisonFailure & 419 \\
java.lang.Exception & 250 \\
java.lang.IllegalStateException & 202 \\
java.lang.NoClassDefFoundError & 197 \\
java.lang.RuntimeException & 191 \\
junit.framework.AssertionFailedError & 163 \\
java.lang.ExceptionInInitializerError & 117 \\
java.io.IOException & 110 \\
Subtotal & \numprint{4452} \\ \hline
Other & \numprint{1928} \\ 
Total & \numprint{6380} \\ \bottomrule
\end{tabular}
\caption{The most common test failure types collected by Repairnator.}
\label{tab:most-common-failures}
\end{table}

\subsection{Recommendations}
We make the following recommendations for anyone aiming at reproducing CI failing builds. 

\subsubsection{The problem of merge commits} 
We have to be careful about the state of a build regarding its related commit, especially in the case of a pull request. 
Each time a commit is created in a pull request, a merge commit is automatically created  in GitHub.
However, GitHub only gives access to the last merge commit of a PR, and not for intermediate commits in the PR.
This means that sometimes, there is a discrepancy between the CI build failure and the actually considered build failure.
We have reported this issue to TravisCI and GitHub.

\textbf{Recommendation \#3}: Take great care of getting the  exact same code state as CI. When getting a build from a pull-request, reproduce yourself the merge commit.

\subsubsection{Managing the dependencies}
When compiling the code, it is very important to properly isolate your dependency manager, so that cached dependencies are not reused from one failure reproduction attempt to another one.
When using Maven, this means using a specific local cache per reproduction attempt.
If not, there is a chance of using another dependency than the one used in the original bug, and hence obtain different spurious behaviors.

\textbf{Recommendation \#4}: Run the bug reproduction (compilation and test execution) in a well-isolated environment. Local caches and containerization help a lot to achieve good isolation.

\section{Patch Synthesis}
\label{sec:stage_repairing}

We discuss in this section the last stage of the Repairnator bot: the patch synthesis itself, i.e. the core program repair algorithm.
We first describe the Repairnator approach, then we present our results, and finally we discuss about our findings.

\subsection{Approach}

We design and implement patch synthesis in three steps.
The first step is to gather information about the project.
The second step consists in launching all automatic repair tools under consideration.
Finally, the third step consists in reporting the resulting patches, if any.

\subsubsection{Gathering project information}

For running the automatic repair tools, we need different pieces of information: 
(1) the location of the source code that is likely to be patched; 
(2) the names of the failing tests; 
(3) the failure type; 
and (4) the location of the compile-time and run-time dependencies of the project to repair. 

The location of the source code is used for synthesizing the patch, and contains file names and line numbers pointing to particular statements. 
It is important to directly give the names of failing tests to the repair tool  in order to save the time of re-executing the entire test suite.
The failure type (e.g. \texttt{NullPointerException} or \texttt{AssertionError}) is used to select the kind of repair tool to use for the patch generation. 
Finally, we need to give the whole dependencies of the project to the repair tool in order to perform any dynamic analysis.

\subsubsection{Launching automatic repair tools}

In Repairnator, We use three different repair tools: NPEFix~\cite{Cornu2015}, Nopol~\cite{Xuan2017}, and Astor~\cite{Martinez2007}.

NPEFix is designed to synthesize patches for \texttt{NullPointer}-\\\texttt{Exception} (NPE) bugs.
Hence, we use it only when a NPE is encountered during the execution of the test suite. 

Astor and Nopol are more generic repair tools. 
Astor~\cite{Martinez2007} is a generate-and-validate repair tool derived from Genprog \cite{genprog}. \\
Nopol~\cite{Xuan2017} is dedicated to repair conditional bugs by modifying if existing conditions or inserting missing preconditions. Repairnator uses the dynamic synthesizer of Nopol \cite{durieux:hal-01279233}.
We use Astor and Nopol in all repair attempts.

\subsubsection{Reporting patches to developer}
\label{sec:reporting-patches}

If a patch is synthesized, the final step is about reporting it to the developers.
In Repairnator, due to the potential presence of ill-formed patches, we always analyze patches before reaching out to the developers.
Thus, we define a Repairnator patch analyst as a member of the Repairnator project who is responsible for performing a sanity check before proposing the patch to the developers of the original project.

When a patch is found, an email is sent to the Repairnator patch analyst. 
The analyst first checks on the GitHub repository of the incriminated project if the contributor has already proposed a patch.
If no patch has been proposed yet, the analyst verifies the bot's patch with the following process.

First, she performs a sanity check to see whether the patch actually fix the failure. Then, she further checks whether the patch is not overfitting~\cite{smith2015cure}: an overfitting patch is a patch that is adequate with respect to the test suite yet incorrect because it is too specific and only fixes the input points of the failing test case but not the whole buggy behavior. 
If the patch is considered as correct by the analyst, she proposes it to the developer (e.g. as a new pull request).

The operation of Repairnator also enables us to collect valuable information for future scientific research onto program repair. 
Consequently, Repairnator also pushes all patches as well as bug reproduction, test failure and repair attempt logs to an archival repository. 
We envision that this data will help software engineering researchers in their future research in the program repair field.

\begin{table}[h]
\centering
\ra{1.3}
\rowcolors{2}{gray!20}{white}
\begin{tabular}{p{30mm}rrr} \toprule
Project & Builds w/ & Nopol & NPEFix \\ 
& patches & patches & patches  \\ \hline
jamesagnew/hapi-fhir & 1 & 35 & 0 \\
spotify/cassandra-reaper & 1 & 1 & 0 \\
xmlunit/xmlunit & 1 & 145 & 0 \\
apache/pdfbox & 1 & 120 & 0 \\
LiveRamp/hank & 1 & 4 & 0 \\
spring-cloud/spring-cloud-dataflow & 1 & 0 & 1 \\
IQSS/dataverse & 2 & 0 & 16 \\
bonigarcia/webdrivermanager & 3 & 30 & 0 \\
GeoWebCache/geowebcache & 1 & 0 & 2 \\
timmolter/XChange & 1 & 0 & 4 \\
phax/jcodemodel & 1 & 624 & 0 \\
phoenixnap/springmvc-raml-plugin & 1 & 348 & 0 \\ \hline
Total & 15 & \numprint{1307} & 23 \\ \bottomrule
\end{tabular}
\caption{Number of builds with at least one test-suite adequate patches.}
\label{tab:build-patches}
\end{table}

\subsection{Results}

We use three different repair tools in Repairnator. However they have not been integrated at the same time: 
Nopol has been used since February 2017;
NPEFix has been used since August 2017; and Astor since September 2017.
Due to the absence of large enough data for Astor, the results and discussion below are exclusively about NPEFix and Nopol.

We present in Table~\ref{tab:build-patches} the test-suite adequate patches for the projects with at least one Nopol or NPEFix patch. 
Recall that a test-suite adequate patch fixes the failing test cases and do not introduce any visible regression.
Moreover a repair tool can synthesize many patches for the same identified bug. 
For example, a precondition in Nopol is synthesized using all available execution context of the failing test: the repair tool may synthesized a \texttt{true} value in the precondition with a code like \texttt{this.equals(this)} but it could also synthesize for the same value: \texttt{this instanceof Object}.
Over the \numprint{3551} successfully reproduced bugs, we collected a total of 15 builds with test-suite adequate patches and \numprint{1307} patches.
We can see that two executions of Nopol in particular produces almost all the patches in proportion. 
This very large number of patches confirms the presence of many tests-suite adequate patches in the repair search space, as pioneeringly shown by Long and Rinard~\cite{Long2016analysis}.

All of those patches were considered as overfitting.
This is a new major piece of evidence, after the recent large scale studies of program repair on real bugs \cite{defects4j-repair}, that overfitting is the main barrier to using program repair in industry.

\begin{lstlisting}[language=diff, caption={Overfitting Repairnator patch for phax/jcodemodel}, label=lst:phax/jcodemodel:potential,float]
@@ JCommentPart.java


- if (aValue == null)
+ if (this.equals((java.lang.Object) this))
    return;
  if (aValue instanceof Object [])
\end{lstlisting}

\begin{lstlisting}[language=diff, caption={The human patch for phax/jcodemodel}, label=lst:phax/jcodemodel:human,float]
@@ JCommentPart.java

   // Only String and AbstractJType are allowed
   if (aValue instanceof String || aValue instanceof AbstractJType) 
     { super.add (aValue); }
-  throw new IllegalArgumentException ("Value is of an unsupported type: " + aValue.getClass ().toString ());
+  else 
+    { throw new IllegalArgumentException ("Value is of an unsupported type: " + aValue.getClass ().toString ()); }
   }
 
\end{lstlisting}

We show in Listing~\ref{lst:phax/jcodemodel:potential} an example of a test-suite adequate yet overfitting patch for project \texttt{phax/jcodemodel}, and in Listing~\ref{lst:phax/jcodemodel:human} the human patch realized for fixing the same bug.
We can see quite confidently that the patch of Listing~\ref{lst:phax/jcodemodel:potential} is overfitting only by reading the proposed condition: the execution of \\ \texttt{this.equals((java.lang.Object) this)} certainly always returns a true value.

\subsection{Recommendations}

\subsubsection{The problem of spurious bugs}
The first pitfall concerning repairing failing builds is about repairing real unexpected behavior and not spurious bugs.
In our context, a spurious bug is a bug that is failing both on CI and locally but for a different root cause.
This is a real problem since it means that Repairnator tries to repair a bug that is different, and a found patch would be irrelevant for the developers.

Currently we cannot ensure that a bug locally re-executed is exactly the same as the one encountered during the TravisCI build. 
Two reasons prevent us to assess this property: the usage of a build script, and the environment of TravisCI.

\paragraph{Build scripts}
Building and testing a code project is not necessarily only about basic compilation and executing test code.
It can involve many more steps: moving resources, processing files, downloading dependencies, etc.
Some of these actions are already automated during the execution of Maven goals, and are then replicated during Repairnator approach.

However developers can significantly modify the default Maven setup or even use an entire build script in TravisCI.
As we do not use those developer provided build scripts, we do not guarantee exactly the same execution condition as the original bug.

\paragraph{TravisCI environment}
TravisCI uses a specific environment for executing the scripts.
This environment can be customized to use, for example, a specific Java version, or even a specific OS (e.g. MacOS or Ubuntu Trusty).

In Repairnator, we execute our build reproduction in the same environment each time, without taking into account this additional environment information, which is sometimes the cause of spurious results in bug reproduction.

\textbf{Recommendation \#5:} Consider engineering the replication of TravisCI environment and run the TravisCI build scripts for repair attempts: the additional effort may be balanced by the number and quality of reproduced failing bugs.

\subsubsection{About multi-module projects}
The second pitfall is related to the design of the repair tools and of the projects to be repaired.
Maven defines a granularity in a project that is called a Maven Module: a module can then be used as a dependency elsewhere in the project, or in another project. 
Hence a Maven project can be a single module, or a multi-module project. 
However a multi-module project means that the source code and the test code of the project will be distributed across multiple directories.

Current repair tools are designed to fix bugs with well-identified location of the test code and the source code.
Multi-module Maven projects do threaten this assumption: some tests might fail because of a piece of code that is in another module of the same project. 
Current repair tools cannot currently repair those bugs spanning multiple modules. 

\textbf{Recommendation \#6:} 
Existing repair tools do not handle
multi-module Maven project. This is major barrier to wide applicability in the field. If you were to design a new repair tool, take care of multi-module projects right at the beginning.

\subsubsection{About Response Time}

The \emph{response time} of a program repair bot is the time duration between the commit date and the patch reporting date.
Currently, the response time is the sum of:
\emph{(1)} T1: the time between commit date and Repairnator execution. In the best case, Repairnator is run by chance just after the commit. In the worst case, Repairnator is run four hours after the commit (because Repairnator is run periodically every four hours).
\emph{(2)} T2: the time to run Repairnator itself 
\emph{(3)} T3: the time for a Repairnator analyst to validate the patch.

While there is no easy answer for reducing T2 and T3, there is one for T1. 
Instead of periodic Repairnator runs, one could plug Repairnator directly in to the CI. 
This is commonly called a \emph{CI hook}: a service called-back for each CI action.
In particular, a CI hook could consist of running Repairnator when a build fails.
Note that, in the context of Repairnator, it is impossible to only use CI hooks instead of periodic runs, because this requires an administrative action from the developers of all \nbprojects under consideration to activate the hook.

\textbf{Recommendation \#7:} Consider implementing CI hooks for program repair bots, it is a good way to minimize the repair bot response time.

\section{Related work}
\label{sec:related}

We present in this section previous works related to software development bots and program repair.

\subsection{Software Development Bots}

Bots are already used by developers, in particular under the form of chatbots. A chatbot is an interactive interface to a system in order to give commands and receive information.
For instance, the notable Hubot project\footnote{https://hubot.GitHub.com} aims at providing a set of APIs for developing chat bots for continuous integration systems. 

The role of bots and how they are improving developer productivity is studied by Storey and Zagalasky~\cite{Storey2016}. They provide some categories to classify the existing development bots using \emph{efficiency} and \emph{effectiveness} criteria. They pinpoint the importance in automating tedious tasks and in keeping developers in the loop by integrating bots in developer existing environment. They also discuss the question of developers trusting bots if they generate artifacts automatically.

A similar problem is studied by Murgia~\emph{et al.}~\cite{murgia2016among}. In this article, they show the compared impacts of two identical bots answering questions of developers on the Stackoverflow platform. The only difference between the two bots is their identity: the first one is presented as a human being, and the other one is clearly displayed (name, avatar) as a programmatic bot. Their results show that the developers had a really higher confidence in results provided by the ``human'' bot. The authors explain that developers certainly have a very low tolerance and very high expectations answers or artifacts generated by bots.

CCBot\cite{carr2017automatic} is a bot dedicated to automatically insert new contracts in C\# projects. It has been created to help developers manage the results of the static analysis tools. The bot is integrated on GitHub and automatically builds projects, analyzes code contracts and proposes code changes for fixing warnings. The code changes are made as pull requests proposed to the developers. CCBot has been validated on 4 C\# projects on GitHub and its authors obtained 22 merged PR.

Balachandran presents ReviewBot~\cite{Balachandran2013b} and its extension called Fix-it~\cite{Balachandran2013}. ``ReviewBot'' is a standalone bot responsible for doing code review of Java programs. The bot is based on static analysis tools to detect standard code violations and common defect patterns. Then ReviewBot has been extended with Fix-it, which aims at automatically fixing some common defects identified during review. Fix-it is based on maintaining an AST of the program and performing AST transformation to fix the bad smells.

Beschastnikh~\emph{et al.} \cite{Beschastnikh2017} set the concept of a common platform for software engineering research tools. They envision in their paper an ecosystem of bots dedicated to software development platform such as GitHub or Bitbucket, which would be able to submit a pull request containing a bug fix, or helping to improve the documentation.

\subsection{Program Repair}
For the high cost of fixing bugs manually, much research effort has been put into the area of automatic program repair in recent years. Automatic program repair aims to automatically eliminate program defects without the intervention of human beings.
We next give an overview of test suite based repair, which is the most widely studied and arguably the standard family of repair techniques. For a complete picture of the field, we refer readers to the paper \cite{repairsurvey}.

Test suite based repair takes as inputs the buggy program and a test suite, which contains some passing tests to specify the desired behaviors of the buggy program and at least one failing test to specify the bug to be repaired, and outputs one or more candidate patches that make all tests pass. In general, test suite based repair techniques first identify the likely buggy statements through fault localization techniques \cite{tarantula,yzxicse,predicateswitching}, and then patch the suspicious statements using certain patch generation strategies. Test suite based repair techniques can further be divided into generate-and-validate and synthesis-based techniques depending on the used patch generation strategy. 

Generate-and-validate repair techniques first search within an established search space to generate a set of candidate patches, and then validate them using the test suite. To establish the search space, proposed strategies in the literature include copying code snippets elsewhere from the same program \cite{genprog} and instantiating human written or learned patch templates \cite{kim2013automatic,spr,gensis}. After establishing the search space, techniques such as genetic programming \cite{genprog} and random search \cite{rsrepair} have been proposed to search the space. 
Synthesis-based techniques first establish repair constraints through the execution of the input test suite, and then get a patch by using program synthesis to solve the constraint. To establish the repair constraint, both symbolic execution \cite{semfix} and concrete execution \cite{Xuan2017} on the test inputs have been employed by existing techniques. Meanwhile, there exist works that try to customize the repair constraint to make the resultant patch more readable \cite{directfix} and increase the scalability of this category of technique \cite{Angelix}. 

Empirical studies have shown that test suite based repair techniques can tackle real faults in real-world programs \cite{genprog, Angelix}. However, an inherent limitation of them is that they use the test suite as the correctness specification to guide the repair process, which is rarely exhaustive in practice. Consequently, the resultant patches can just overfit to the used tests but break untested but desired functionality. Indeed, it has been shown that overfitting is a serious issue associated with test suite based repair techniques \cite{qi2015efficient,smith2015cure,defects4j-repair}. Given the seriousness of the overfitting problem, several recent studies \cite{yzx,qixinISSTA,yangFSE} have tried to employ test case generation to alleviate the issue.

\section{Conclusion}
\label{sec:conclusion}

Software development is evolving very fast: today, almost all developers are using continuous integration to assess code quality and speed up deployment.
Tomorrow we envision that software repair bots will be a standard tool for helping developers to maintain large code bases. 

We have built and designed a software development bot dedicated to program repair.
The bot has been used for 11 months, has managed to reproduce a large set of failing CI builds, and generated patches for 15 builds.
In this paper, we share this unique experience through 7 recommendations in order to help future developers to design and operate their own repair bots.

Repairnator has not yet succeeded in proposing an effective patch to a human developer.
However, Repairnator is already a success, it has enabled us to collect unique empirical data on the key challenges of program repair. 
This data will help the research community to tackle those hard problems and will contribute to eventually achieve true industrial application of program repair.

\section*{Acknowledgement}

This research has been supported by the InriaHub program and the Wallenberg Autonomous Systems and Software Program (WASP).

\bibliographystyle{abbrv}

\bibliography{repairnator-SEIP} 
\end{document}